\documentclass[10pt]{amsart}
\usepackage{amsmath}
\usepackage{amssymb}
\usepackage{amsfonts}
\usepackage{amsthm}
\usepackage{amsxtra}
\usepackage{float}
\usepackage[all,web,arc,poly,dvips,color]{xy}
\UseCrayolaColors
\usepackage{epsfig}
\usepackage[dvips]{color}
\usepackage{array}
\usepackage{pifont}
\usepackage{multirow}
\usepackage{graphics}

\theoremstyle{plain}

\newtheorem{corollary}{Corollary}

\newtheorem{proposition}{Proposition}

\theoremstyle{definition}

\theoremstyle{remark}

\newcommand{\C}{\mathbb C}

\newcommand{\Z}{\mathbb Z}
\newcommand{\N}{\mathbb N}
\newcommand{\T}{\mathbb T}

\newcommand{\DySt}{\displaystyle}

\newcommand{\PVI}{${\rm P}_{\rm VI}\;$}

\newcommand{\half}{
        {\lower0.00ex\hbox{\raise.6ex\hbox{\the\scriptfont0 1}
                           \kern-.5em\slash\kern-.1em\lower.45ex
                                     \hbox{\the\scriptfont0 2}}}}
 \newcommand{\thalf}{
        {\lower0.00ex\hbox{\raise.6ex\hbox{\the\scriptfont0 3}
                           \kern-.5em\slash\kern-.1em\lower.45ex
                                     \hbox{\the\scriptfont0 2}}}}                             \newcommand{\fhalf}{
        {\lower0.00ex\hbox{\raise.6ex\hbox{\the\scriptfont0 5}
                           \kern-.5em\slash\kern-.1em\lower.45ex
                                     \hbox{\the\scriptfont0 2}}}}       
\newcommand{\quarter}{
        {\lower0.00ex\hbox{\raise.6ex\hbox{\the\scriptfont0 1}
                           \kern-.5em\slash\kern-.1em\lower.45ex
                                     \hbox{\the\scriptfont0 4}}}}
\newcommand{\tquarter}{
        {\lower0.00ex\hbox{\raise.6ex\hbox{\the\scriptfont0 3}
                           \kern-.5em\slash\kern-.1em\lower.45ex
                                     \hbox{\the\scriptfont0 4}}}}
\newcommand{\eighth}{
        {\lower0.00ex\hbox{\raise.6ex\hbox{\the\scriptfont0 1}
                           \kern-.5em\slash\kern-.1em\lower.45ex
                                     \hbox{\the\scriptfont0 8}}}}
\newcommand{\teighths}{
        {\lower0.00ex\hbox{\raise.6ex\hbox{\the\scriptfont0 3}
                           \kern-.5em\slash\kern-.1em\lower.45ex
                                     \hbox{\the\scriptfont0 8}}}}       
\newcommand{\othird}{
        {\lower0.00ex\hbox{\raise.6ex\hbox{\the\scriptfont0 1}
                           \kern-.5em\slash\kern-.1em\lower.45ex
                                     \hbox{\the\scriptfont0 3}}}}

\begin{document}

\title[]
{Isomonodromic deformation theory and the next-to-diagonal correlations of the anisotropic square lattice Ising model.}

\author{N.S.~Witte}
\address{Department of Mathematics and Statistics,
University of Melbourne,Victoria 3010, Australia}
\email{\tt n.witte@ms.unimelb.edu.au}

\begin{abstract}
In 1980 Jimbo and Miwa evaluated the diagonal two-point correlation function 
of the square lattice Ising model as a $\tau$-function of the sixth Painlev\'e 
system by constructing an associated isomonodromic system within their theory 
of holonomic quantum fields. More recently an alternative isomonodromy theory 
was constructed based on bi-orthogonal polynomials on the unit circle with 
regular semi-classical weights, for which the diagonal Ising correlations 
arise as the leading coefficient of the polynomials specialised appropriately. 
Here we demonstrate that the  next-to-diagonal correlations of the anisotropic 
Ising model are evaluated as one of the elements of this isomonodromic system 
or essentially as the Cauchy-Hilbert transform of one of the bi-orthogonal 
polynomials.
\end{abstract}

\subjclass[2000]{82B20,34M55,33C45}
\maketitle

For the square lattice Ising model on the infinite lattice an unpublished result 
of Onsager (see \cite{McCW_1973}) gives that the diagonal spin-spin correlation
$ \langle \sigma_{0,0}\sigma_{N,N} \rangle $ has the Toeplitz determinant form
\begin{equation}
   \langle \sigma_{0,0}\sigma_{N,N} \rangle = 
   \det (a_{i-j}(k))_{1 \leq i,j \leq N} ,
\label{IM_ssDiag}
\end{equation}
where the elements are given by 
\begin{equation}
 a_n = \int^{\pi}_{-\pi}\frac{d\theta}{2\pi}
   \frac{k\cos n\theta-\cos(n\!-\!1)\theta}{\sqrt{k^2+1-2k\cos\theta}} .                        
\end{equation}
A significant development occurred when Jimbo and Miwa \cite{JM_1980,JM_1980errata} 
identified 
(\ref{IM_ssDiag}) as the $ \tau$-function of a \PVI system. This identification 
has the consequence of allowing (\ref{IM_ssDiag}) to be characterised in terms of 
a solution of the $ \sigma$-form of the Painlev\'e VI equation, a second order
second degree ordinary differential equation with respect to $ t:=k^{\pm 2} $ with 
parameter $ N $, or as the solution of coupled recurrence 
relations in $ N $ with parameter $ t $, which were subsequently shown to be 
equivalent to the discrete Painlev\'e V equation.
This was derived from the monodromy preserving deformation of a certain linear 
system as a particular example of their general theory of holonomic quantum fields 
\cite{MJ_1982,SMJ_1980,KK_1980},
however the theoretical machinery employed there was never put to use on related 
problems arising from the Ising model. See the forthcoming monograph \cite{Pa_2007} 
on recent progress utilising this viewpoint. In a recent work \cite{FW_2004} 
Forrester and the present author identified (\ref{IM_ssDiag}) as a $ \tau$-function 
in the Okamoto theory of \PVI \cite{Ok_1987a} and subsequently developed an 
alternative isomonodromic theory \cite{FW_2004a} founded on bi-orthogonal systems 
on the unit circle with regular semi-classical weights. We remark that a result of 
Borodin \cite{Bo_2001} can also be used for the same purpose.

In a further development Au-Yang and Perk \cite{A-YP_1987},\cite{A-YP_2001a} 
discovered that the next-to-diagonal spin-spin correlations have the bordered 
Toeplitz determinant form
\begin{equation}
\langle \sigma_{0,0}\sigma_{N,N-1} \rangle
  = \det\begin{pmatrix}
          a_{0}  & \cdots & a_{-N+2} & b_{N-1} \cr
          a_{1} & \cdots & a_{-N+3} & b_{N-2} \cr
          \vdots & \vdots & \vdots  & \vdots  \cr
          a_{N-1} & \cdots & a_{1}  & b_{0}   \cr
        \end{pmatrix} ,\; N \geq 1
\label{IM_nextDiag:a}
\end{equation}
where the elements $ a_n $ are the same as those above and the $ b_n $ are given by
\begin{equation}
 b_n = \int^{\pi}_{-\pi}\frac{d\theta}{2\pi}\frac{\bar{C}}{\sqrt{k^2+1-2k\cos\theta}}
     \frac{(k\bar{S}-S)\cos n\theta+kS\cos(n\!-\!1)\theta-\bar{S}\cos(n\!+\!1)\theta}
           {S^2+\bar{S}^2+2k\cos\theta} ,
\label{IM_nextDiag:bD}
\end{equation}
(the definitions of the model parameters $ k,S,\bar{S} $ are given in the following paragraph). 
The task of the present study is to answer the following questions - can this 
correlation be evaluated in terms of a Painlev\'e-type function and if so which one?
The answer is in the affirmative and we identify the function in Proposition 
\ref{next-diag-Corr}. 
In order to understand the result for the next-to-diagonal correlations in its 
proper context we will need to revise some relevant known results for the diagonal
correlations. In fact even in an algorithmic sense in order to compute the
next-to-diagonal correlations one has to first compute the diagonal ones.

Consider the Ising model with spins $ \sigma_r \in \{-1,1\} $ located at site $ r=(i,j) $ 
on a square lattice of dimension $(2L+1) \times (2L+1)$, centred about the origin.
The first co-ordinate of a site refers to the horizontal or $ x $-direction and the 
second to the vertical or $ y $-direction, which is the convention opposite to that
of McCoy and Wu \cite{McCW_1973} and early studies where the first co-ordinate
labeled the rows in ascending order and the second the columns from left to right.
We will focus on the homogeneous but anisotropic Ising model where  
the dimensionless nearest neighbour couplings are equal to $ \bar{K} $ and $ K $ in 
the $ x $ and $ y $ directions respectively (see e.g. \cite{Ba_1982}). 

\begin{figure}[H]
\[
 \begin{xy}
    *\xybox{0;<2cm,0cm>:<0cm,2cm>::
	   ,0="O"*{\bullet}*!RU{(i,j)\;}
	   ,{\ar@{-}^{\DySt \bar{K}} "O";"O"+<2cm,0cm>}
	   ,{\ar@{-}^{\DySt K} "O";"O"+<0cm,2cm>}
     ,"O";"O"+<-2cm,0cm>**@{-}
     ,"O";"O"+<0cm,-2cm>**@{-}
     ,"O"+(0,1)="U"*{\bullet}*!RD{(i,j\!+\!1)\;}
     ,"U";"U"+<2cm,0cm>**@{-}
     ,"U";"U"+<-2cm,0cm>**@{-}
     ,"U";"U"+<0cm,2cm>**@{-}
     ,"U"+(1,0)="D"*{\bullet}*!LD{\;(i\!+\!1,j\!+\!1)}
     ,"D";"D"+<2cm,0cm>**@{-}
     ,"D";"D"+<0cm,-2cm>**@{-}
     ,"D";"D"+<0cm,2cm>**@{-}
     ,"O"+(1,0)="R"*{\bullet}*!LU{\;(i\!+\!1,j)}
     ,"R";"R"+<2cm,0cm>**@{-}
     ,"R";"R"+<0cm,-2cm>**@{-}
           }  
 \end{xy} 
\]
\caption{Co-ordinate system and couplings for the homogeneous anisotropic square lattice Ising model}
\end{figure}
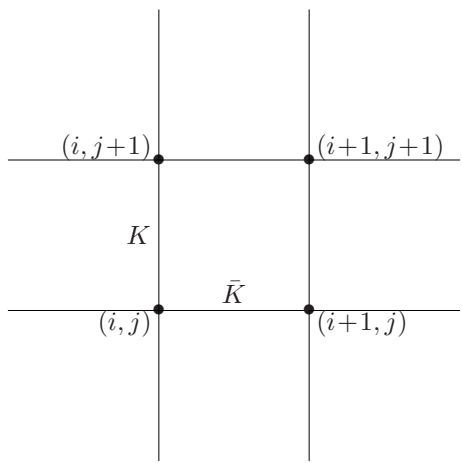

The probability density function for configuration $ \{\sigma_{ij}\}_{i,j=-L}^{L} $ 
is given by
\begin{equation}
 {\rm Pr}(\{\sigma_{ij}\}_{i,j=-L}^{L}) = \frac{1}{Z_{2L+1}}
     \exp \Big[ \bar{K} \sum_{j=-L}^L \sum_{i=-L}^{L-1}\sigma_{ij}\sigma_{i+1\, j}
               + K \sum_{i=-L}^L \sum_{j=-L}^{L-1}\sigma_{ij}\sigma_{i\, j+1} \Big],
\end{equation}
and averages are defined by
\begin{equation}
  \langle \cdot \rangle
  = \sum_{\sigma_{ij}} \cdot\;{\rm Pr}(\{\sigma_{ij}\}_{i,j=-L}^{L}).
\end{equation}
The normalisation $ Z_{2L+1} $ is the partition function and conventionally periodic
boundary conditions, $ \sigma_{i,L+1}=\sigma_{i,-L} $, $ \sigma_{L+1,j}=\sigma_{-L,j} $
for all $ i,j $, are taken for convenience.
In all such averages the thermodynamic limit is taken
$ \lim_{L\to \infty}\langle \cdot \rangle $ keeping $ K,\bar{K} $ fixed.
The relevant variables in our study are the following variables 
$ k,S,\bar{S},C,\bar{C} $ defined by 
\begin{equation}
 S:=\sinh 2K, \quad \bar{S}:=\sinh2\bar{K}, \quad C:=\cosh 2K, \quad \bar{C}:=\cosh 2\bar{K},
 \quad k:=S\bar{S}
\label{ising_param}
\end{equation}
We will only treat the system in the ferromagnetic regime $ K,\bar{K}> 0 $ and 
$ k\in (0,\infty) $, which exhibits a phase transition at the critical value $ k=1 $.
We will find subsequently that, from the point of view of the theory of
isomonodromic systems, that the next-to-diagonal correlations are functions of the 
two complex variables, $ k $ and one of $ S,\bar{S} $, with $ k $ playing the role 
of the deformation variable and $ -\bar{S}/S $ the spectral variable. While all 
of the results can be continued into the complex plane $ k,S \in \C $ suitably 
restricted we may often only state them for the physical regime $ k,S,\bar{S} \in (0,\infty) $. 
Corresponding to the Ising model is a dual partner Ising model, which is related 
to the original by the duality transformation or involution
\begin{gather}
  \sigma_r \mapsto \mu_r, \quad
  \langle \sigma_{r_1} \ldots \sigma_{r_n} \rangle \mapsto 
  \langle \mu_{r_1} \ldots \mu_{r_n} \rangle,
  \label{dual:a} \\
  k \mapsto \frac{1}{k}, \quad
  S \mapsto \frac{1}{\bar{S}}, \quad
  \bar{S} \mapsto \frac{1}{S} .
  \label{dual:b}
\end{gather}
The dynamic variables $ \mu_r $ are known as the disorder variables and can be 
given an interpretation in terms of the spins variables $ \sigma_r $ \cite{KK_1980}.

The appearance of {\em Toeplitz} determinants such as those of (\ref{IM_ssDiag}) is 
indicative of several structures and the most general of these is averages
over the unitary group. Let $ U\in U(N) $ have eigenvalues 
$ z_1=e^{i \theta_1}, \dots, z_N=e^{i \theta_N} $. 
The unitary group $U(N)$ with Haar (uniform) measure has eigenvalue probability 
density function 
\begin{equation}
   \frac{1}{(2 \pi )^N N!} \prod_{1 \le j < k \le N} | z_k - z_j |^2,
   \quad z_l := e^{i \theta_l} \in \T, \quad \theta_l \in (-\pi,\pi] ,
\label{ops_Haar}
\end{equation}
with respect to Lebesgue measure $ d\theta_1\cdots d\theta_N $ (see e.g.~\cite[Chapter 2]{rmt_Fo})
where $ \T = \{z \in \C: |z|=1 \} $. A well known identity \cite{ops_Sz} relates 
averages of class functions, in particular products of a function $ w(z) $ over
the eigenvalues, to the Toeplitz determinant
\begin{equation}
  I^{\epsilon}_N[w] := 
  \Big \langle \prod_{l=1}^N w(z_l)z^{\epsilon}_l \Big \rangle_{U(N)} =
  \det[ w_{-\epsilon+j-k} ]_{j,k=1,\dots,N}, \quad \epsilon\in \Z,\; N\geq 1 .
\label{ops_Uavge}
\end{equation}
By convention we set $ I^{\epsilon}_0=1 $ and use the short-hand notation
$ I_N:=I^{0}_N $.
We identify $ w(z) $ as a weight function with the Fourier decomposition
\begin{equation}
   w(z) = \sum_{l\in\Z} w_l z^l .
\end{equation}

The specific Fourier coefficients appearing in the diagonal Ising correlations 
(\ref{IM_ssDiag}) are
\begin{equation}
 a_n(k)
  = \int_{\T}\frac{d\zeta}{2\pi i\zeta}\zeta^{n}
                           \sqrt{\frac{1-k^{-1}\zeta^{-1}}{1-k^{-1}\zeta}}
  =  \int^{\pi}_{-\pi} \frac{d\theta}{2\pi}
   \frac{k\cos n\theta-\cos(n-1)\theta}{\sqrt{k^2+1-2k\cos\theta}} .
\label{IM_ssSymbol}                           
\end{equation}
The implied weight is
\begin{equation}
   a(\zeta;k) = \begin{cases}
    k^{-1/2}\zeta^{1/2}(\zeta-k^{-1})^{-1/2}(k-\zeta)^{1/2}, & 1<k<\infty \\
   -k^{-1/2}\zeta^{1/2}(k^{-1}-\zeta)^{-1/2}(\zeta-k)^{1/2}, & 0\leq k<1
                \end{cases} .
\label{VI_wgt:a}
\end{equation}
The analytic structure is different depending on $ k > 1 $ (low temperature phase) 
or $ k < 1 $ (high temperature phase). The reason for the phase change of 
$ e^{-\pi i} $ in the weight is because of the argument changes 
\begin{equation}
  \zeta-k = e^{-\pi i}(k-\zeta), \qquad 
  k^{-1}-\zeta = e^{\pi i}(\zeta-k^{-1}) ,
\end{equation}
as $ k $ goes from the $ k>1 $ to the $ k<1 $ regime.
The correlation function for the disorder variables is
\begin{equation}
\langle \mu_{0,0}\mu_{N,N} \rangle
  =    \det (\tilde{a}_{i-j}(k))_{1 \leq i,j \leq N} ,
\end{equation}
where
\begin{equation}
 \tilde{a}_n(k)
  = \int_{\T}\frac{d\zeta}{2\pi i\zeta}\zeta^{n}
                           \sqrt{\frac{1-k\zeta^{-1}}{1-k\zeta}}
  =  \int^{\pi}_{-\pi} \frac{d\theta}{2\pi}
   \frac{\cos n\theta-k\cos(n-1)\theta}{\sqrt{k^2+1-2k\cos\theta}} .
\end{equation}
The weight is
\begin{equation}
   \tilde{a}(\zeta;k) 
    =  \begin{cases}
   -k^{1/2}\zeta^{1/2}(k-\zeta)^{-1/2}(\zeta-k^{-1})^{1/2}, & 1<k<\infty \\
    k^{1/2}\zeta^{1/2}(\zeta-k)^{-1/2}(k^{-1}-\zeta)^{1/2}, & 0\leq k<1
       \end{cases}
\label{VI_wgt:b}
\end{equation}
Although we use the same notation for the Toeplitz elements as Au-Yang and Perk \cite{A-YP_2001a} 
the relationship between our elements and theirs is $ a_n = a^{A-YP}_{-n} $ and 
$ \tilde{a}_n = \tilde{a}^{A-YP}_{-n} $. 

From the viewpoint of the work \cite{FW_2004a} the weights (\ref{VI_wgt:a}), 
(\ref{VI_wgt:b}) are particular examples of the {\em regular semi-classical class}, 
characterised by a special structure of their logarithmic derivatives
\begin{equation}
  \frac{1}{w(z)}\frac{d}{dz}w(z) := \frac{2V(z)}{W(z)}
  = \sum^3_{j=1}\frac{\rho_j}{z-z_j}, \quad \rho_j \in \C.
\label{ops_scwgt2}
\end{equation}
Here $ V(z) $, $ W(z) $ are polynomials with $ {\rm deg}V(z) < 3, {\rm deg}W(z)=3 $.
The data for the weight (\ref{VI_wgt:a}) is then
\begin{equation}
  \{z_j\}^{3}_{j=1} = \{0,k^{-1},k\}, \quad
  \{\rho_j\}^{3}_{j=1} = \{\half,-\half,\half\} .
\end{equation}
The data for the other weight is (\ref{VI_wgt:b})
\begin{equation}
  \{z_j\}^{3}_{j=1} = \{0,k,k^{-1}\}, \quad
  \{\rho_j\}^{3}_{j=1} = \{\half,-\half,\half\} ,
\end{equation}
which is the same as the previous case except for the exchange in the position of
variable singularities. A particular observation in the Ising model case is that the
Toeplitz matrix is not hermitian and the weight $ w(z) $ is complex for real and 
physical $ k \in (0,\infty) $, $ z\in \T $. The duality transformation is simply a 
transposition of the singular points $ z_2 \leftrightarrow z_3 $ and at the critical 
temperature these two singularities coalesce.

An important identity relating the dual Toeplitz elements to the direct ones
is the following well known duality relation \cite{A-YP_2001a}.
\begin{proposition}\label{IM_dual.1}
For all $ k $ and $ n $ we have
 \begin{equation}
   \tilde{a}_{n}(k) = a_{n}(k^{-1}) = -a_{-n+1}(k) .
\end{equation}
The two weights are related by the duality transformation
\begin{equation}
  \tilde{a}(\zeta;k) = a(\zeta;k^{-1}).
\end{equation}
\end{proposition}

By regarding the Fourier integral in (\ref{IM_ssSymbol}) as a contour integral,
and changing the contour of integration one obtains
the well known fact that the Toeplitz elements in the low temperature regime are 
given by
\begin{align}
  a_{n}  & = -\frac{\Gamma(n-\half)\Gamma(\thalf)}{\pi\Gamma(n+1)}k^{-n}
             {}_2F_1(\half,n-\half;n+1;k^{-2}), \quad n \geq 0,
         \label{IM_Toep:a}\\
  a_{-n} & = \frac{\Gamma(n+\half)\Gamma(\half)}{\pi\Gamma(n+1)}k^{-n}
             {}_2F_1(-\half,n+\half;n+1;k^{-2}), \quad n \geq 0,
         \label{IM_Toep:b}
\end{align}
whilst those in the high temperature regime are
\begin{align}
  a_{n}  & = -\frac{\Gamma(n-\half)\Gamma(\half)}{\pi\Gamma(n)}k^{n-1}
             {}_2F_1(-\half,n-\half;n;k^{2}), \quad n \geq 1.
         \label{IM_Toep:c}\\
  a_{-n} & = \frac{\Gamma(n+\half)\Gamma(\thalf)}{\pi\Gamma(n+2)}k^{n+1}
             {}_2F_1(\half,n+\half;n+2;k^{2}), \quad n \geq -1,
         \label{IM_Toep:d}
\end{align}
These elements are expressible as linear combinations of the complete first and
second elliptic integrals $ {\rm K} $, $ {\rm E} $ with arguments $ k^{-1} $ and
$ k $ respectively \cite{GS_1984} and with coefficients polynomial in these arguments. 
In the ensuing discussion we adopt the following shorthand notation for the
complete elliptic integrals of the first kind
\begin{equation}
  {\rm K}_{<} := {\rm K}(k),\; {\rm K}_{>} := {\rm K}(k^{-1}),\;
  {\rm K}_{\lozenge} := {\rm K}(k_{\lozenge}),
\end{equation}
with analogous notation for the second kind and 
where $ k_{\lozenge} = 2\sqrt{k}/(k+1) $ is the inverse Landen transformation.
The complementary modulus is defined $ k'_{\lozenge}:=\sqrt{1-k^2_{\lozenge}} $.

The complex weight $ w(z) $ with support contained in $ \T $ implicitly defines a 
system of {\em bi-orthogonal polynomials} 
$ \{\phi_n(z),\bar{\phi}_n(z)\}^{\infty}_{n=0} $ on the unit circle by the 
orthogonality relation
\begin{equation}
  \int_{\T} \frac{d\zeta}{2\pi i\zeta} w(\zeta)\phi_m(\zeta)\bar{\phi}_n(\bar{\zeta})
   = \delta_{m,n} ,
\label{ops_onorm}
\end{equation}
whose existence is ensured if and only if $ I_n \neq 0 $ for all $ n\in\N $.
Notwithstanding the notation, $ \bar{\phi}_n $ is not in general equal to the 
complex conjugate of $ \phi_n $ and independent of it. The leading and trailing 
coefficients of these polynomials
\begin{align}
  \phi_n(z)       &= \kappa_nz^n + \ldots +\phi_n(0) ,
  \nonumber\\
  \bar{\phi}_n(z) &= \kappa_nz^n + \ldots +\bar{\phi}_n(0) ,
  \nonumber
\end{align}
occupy an important role in the theory where again $ \bar{\phi}_n(0) $ are not in 
general equal to the corresponding complex conjugate. With the so-called reflection 
or Verblunsky coefficients specified by
\begin{equation}
  r_n := \frac{\phi_n(0)}{\kappa_n}, \quad
  \bar{r}_n := \frac{\bar{\phi}_n(0)}{\kappa_n} ,
\end{equation}
it is a well known result in the theory of Toeplitz determinants that
\begin{equation}
   \frac{I_{n+1}[w]I_{n-1}[w]}{(I_{n}[w])^2}
   = 1 - r_{n}\bar{r}_n, \quad
  \kappa^2_n = \frac{I_{n}}{I_{n+1}}, \quad n\geq 1 .
\label{ops_I0}
\end{equation}
Rather than dealing with $ \bar{\phi}_n $ it is advantageous to define the reciprocal 
polynomial $ \phi^*_n(z) $ by
\begin{equation}
  \phi^*_n(z) := z^n\bar{\phi}_n(1/z) .
\label{ops_recip}
\end{equation}
In addition to the polynomial pair $ \phi_n $, $ \phi^*_n $ we require two 
non-polynomial solutions of the fundamental recurrence relations appearing in the
theory \cite{FW_2004a},
\begin{align}
   \epsilon_n(z)
   &:= \int_{\T}\frac{d\zeta}{2\pi i\zeta}\frac{\zeta+z}{\zeta-z}w(\zeta)
                   \phi_n(\zeta) ,\quad n\geq 1,
   \label{ops_eps:a} \\
   \epsilon^*_n(z)
   &:= \frac{1}{\kappa_n} 
          -\int_{\T}\frac{d\zeta}{2\pi i\zeta}\frac{\zeta+z}{\zeta-z}w(\zeta)
                   \phi^*_n(\zeta) ,\quad n\geq 1 .
   \label{ops_eps:b}
\end{align}
These form a matrix system 
\begin{equation}
   Y_n(z;t) :=
   \begin{pmatrix}
          \phi_n(z)   &  \epsilon_n(z)/w(z) \cr
          \phi^*_n(z) & -\epsilon^*_n(z)/w(z) \cr
   \end{pmatrix} ,
\label{ops_Ydefn}
\end{equation}
which, for regular semi-classical weights, has the property \cite{FW_2004a} that 
their 
monodromy data in the complex spectral $ z $-plane is preserved under arbitrary
deformations of the singularities $ z_j $.

From the Toeplitz determinant formula (\ref{IM_ssDiag}) we observe that
\begin{equation}
  \langle \sigma_{0,0}\sigma_{N,N} \rangle
    = \det[a_{j-k}]_{j,k=0,\ldots,N-1} = I_N[a(\zeta;k)] := I_N(k) ,
  \label{Diag:a}
\end{equation}
and apply the known results of Subsection 3.1 in \cite{FW_2004b} which provides 
the following recurrence scheme for the diagonal correlations.

\begin{corollary}[\cite{FW_2004b}]\label{IM_recur}
The diagonal correlation function for the Ising model valid in both the low and
high temperature phases for $ N \geq 1 $ is determined by
\begin{equation}
 \frac{\langle \sigma_{0,0}\sigma_{N+1,N+1} \rangle\langle \sigma_{0,0}\sigma_{N-1,N-1} \rangle}
      {\langle \sigma_{0,0}\sigma_{N,N} \rangle^2}
  = 1-r_{N}\bar{r}_{N},
\end{equation}
along with the quasi-linear $ 2/1 $ 
\begin{multline}
   (2N+3)(1-r_{N}\bar{r}_{N})r_{N+1}
     - 2N\left[ k+k^{-1}+(2N-1)r_{N}\bar{r}_{N-1} \right]r_{N} \\
     + (2N-3)\left[ (2N-1)r_{N}\bar{r}_{N}+1 \right]r_{N-1} = 0,
   \label{ising_rRecur:a}
\end{multline}
and $ 1/2 $ recurrence relation
\begin{multline}
   (2N+1)(1-r_{N}\bar{r}_{N})\bar{r}_{N+1}
     - 2N\left[ k+k^{-1}-(2N-3)\bar{r}_{N}r_{N-1} \right]\bar{r}_{N} \\
     + (2N-1)\left[ -(2N+1)r_{N}\bar{r}_{N}+1 \right]\bar{r}_{N-1} = 0,
   \label{ising_rRecur:b}
\end{multline}
subject to initial conditions $ r_{0} = \bar{r}_{0} = 1 $ and 
\begin{gather}
   r_{1} = \begin{cases}
   \frac{\DySt k^2-2}{\DySt 3k}
        +\frac{\DySt 1-k^2}{\DySt 3k}\frac{\DySt {\rm K}_{>}}{\DySt {\rm E}_{>}}, & 1<k<\infty \\
   \frac{\DySt 1}{\DySt 3}\left[
        -\frac{\DySt 2}{\DySt k} + \frac{\DySt k{\rm E}_{<}}{\DySt (k^2-1){\rm K}_{<}+{\rm E}_{<}}
                          \right], & 0\leq k<1
           \end{cases} , 
   \\
   = \frac{1}{3} \left[ -2\frac{1+k'_{\lozenge}}{1-k'_{\lozenge}} 
                        +\frac{1-k'_{\lozenge}}{1+k'_{\lozenge}}
                         \frac{{\rm E}_{\lozenge}+k'_{\lozenge}{\rm K}_{\lozenge}}{{\rm E}_{\lozenge}-k'_{\lozenge}{\rm K}_{\lozenge}}
                 \right] ,
   \\
   \bar{r}_{1} = \begin{cases}
   k+\frac{\DySt 1-k^2}{\DySt k}\frac{\DySt {\rm K}_{>}}{\DySt {\rm E}_{>}}, & 1<k<\infty \\
   \frac{\DySt k{\rm E}_{<}}{\DySt (k^2-1){\rm K}_{<}+{\rm E}_{<}}, & 0\leq k<1
                 \end{cases} ,
   \\
   =  \frac{1-k'_{\lozenge}}{1+k'_{\lozenge}}
                         \frac{{\rm E}_{\lozenge}+k'_{\lozenge}{\rm K}_{\lozenge}}{{\rm E}_{\lozenge}-k'_{\lozenge}{\rm K}_{\lozenge}} .
\end{gather}
The initial values of the correlations are 
\begin{gather}
  \langle \sigma_{0,0}\sigma_{1,1} \rangle
    = a_{0}
    = \begin{cases}
        \frac{\DySt 2}{\DySt \pi}{\rm E}_{>}, & 1<k<\infty \\
        \frac{\DySt 2}{\DySt \pi k}
        \left[(k^2-1){\rm K}_{<}+{\rm E}_{<}\right], & 0\leq k<1
      \end{cases} \\
    = \frac{2}{\pi}\frac{1}{1-k'_{\lozenge}}\left[ {\rm E}_{\lozenge}-k'_{\lozenge}{\rm K}_{\lozenge} \right] .
\label{ising_rRecur:c}
\end{gather}
\end{corollary}

A consequence of the duality relation (\ref{IM_dual.1}) are the following
obvious relations amongst the coefficients of the bi-orthogonal polynomial system.
\begin{proposition}
For all $ n $ and $ k $ we have
\begin{align}
  I^{\varepsilon}_n[\tilde{a}] & = (-1)^nI^{-1-\varepsilon}_n[a] \\
  \bar{r}_n[\tilde{a}] & = \frac{1}{\bar{r}_n[a]}
\end{align}
\end {proposition} 

Now we turn our attention to the object of the present study - the evaluation
of the next-to-diagonal correlations.
Let us recall that the elements $ b_n $ of the bordered Toeplitz determinant (\ref{IM_nextDiag:bD}) 
can be written as
\begin{equation}
 b_n = \bar{C} \int_{\T}\frac{d\zeta}{2\pi i}\frac{\zeta^{n}}{\bar{S}+S\zeta}
                           \sqrt{\frac{k/\zeta-1}{k\zeta-1}} .
\label{IM_nextDiag:b}
\end{equation}
These elements will also have complete elliptic function representations
however for the anisotropic model we require the complete third elliptic integral 
defined by
\begin{equation}
   \Pi(n,k) := \int^{\pi/2}_{0}\frac{d\phi}{\sqrt{1-k^2\sin^2\phi}}
                               \frac{1}{1-n\sin^2\phi} .
\end{equation}
We also adopt a notational shorthand for these, analogous to that for the 
first and second integrals
\begin{equation}
   \Pi_{<} := \Pi(-S^2,k),\; \Pi_{>} := \Pi(-1/\bar{S}^2,k^{-1}),\;
   \Pi_{\lozenge} := \Pi(-4k(\bar{S}-S)^{-2},k_{\lozenge}) .
\end{equation}
We note that $ \Pi_{\lozenge} $ is not analytic at $ \bar{S}=S $ and in fact 
has a discontinuity there of the following form
\begin{equation}
   \Pi_{\lozenge} = \frac{\pi}{2}{\rm sgn}(\bar{S}-S) + {\rm O}(\bar{S}-S),
   \quad\text{as $ \bar{S} \to S $.}
\label{Pi_discont}
\end{equation}
The first correlation in this sequence ($ N=1 $) has the elliptic function 
evaluation
\begin{gather}
  \langle \sigma_{0,0}\sigma_{1,0} \rangle = b_0
    = \begin{cases}
       \frac{\DySt 2\bar{C}}{\DySt \pi kS} \vphantom{\bigg(}
       \left[ C^2\Pi_{>}-{\rm K}_{>} \right], & 1<k<\infty \\
       \frac{\DySt 2\bar{C}}{\DySt \pi S} \vphantom{\bigg(}
       \left[ C^2\Pi_{<}-{\rm K}_{<} \right], & 0\leq k<1
      \end{cases} ,
\label{IM_nD:1} \\
    = \frac{\bar{C}(1+k'_{\lozenge})}{2\pi S}\left[ 
            C^2\frac{\bar{S}+S}{\bar{S}-S}\Pi_{\lozenge}
           +(S^2-1){\rm K}_{\lozenge} \right] + \frac{C}{S}\Theta(S-\bar{S}), \; 0\leq k<\infty ,
\label{IM_nD:2}
\end{gather}
where $ \Theta(x) $ is the Heaviside step function. The term with the step function in 
(\ref{IM_nD:2}) is necessary to compensate for the discontinuity in $ \Pi_{\lozenge} $ as
given in (\ref{Pi_discont}) in order that the correlation function remain continuous 
at $ \bar{S}=S $. The second correlation function ($ N=2 $) has the evaluation
\begin{gather}
  \langle \sigma_{0,0}\sigma_{2,1} \rangle = 
  \nonumber \\
    \begin{cases}
     \frac{\DySt 4\bar{C}}{\DySt \pi^2 k^3S} \vphantom{\bigg(} \Big\{ 
          C^2\left[k^2(1-\bar{S}^2){\rm E}_{>}+(k^2-1)\bar{S}^2{\rm K}_{>}\right]
             \Pi_{>} \\
          \phantom{\frac{\DySt 4\bar{C}}{\DySt \pi^2 k^3S}\Big\{}
          +k^4{\rm E}_{>}^2+(1-k^2)\bar{S}^2{\rm K}_{>}^2+
           k^2(\bar{S}^2-k^2){\rm E}_{>}{\rm K}_{>} \Big\}, & 1<k<\infty
     \\
     \frac{\DySt 4\bar{C}}{\DySt \pi^2 kS} \vphantom{\bigg(} \Big\{ 
          C^2\left[(k^2-1){\rm K}_{<}+(1-\bar{S}^2){\rm E}_{<}\right]\Pi_{<} \\
           \phantom{\frac{\DySt 4\bar{C}}{\DySt \pi^2 kS}\Big\{}
          +{\rm E}_{<}^2+(1-k^2){\rm K}_{<}^2+
           (C^2\bar{S}^2-2){\rm E}_{<}{\rm K}_{<} \Big\}, & 0\leq k<1
    \end{cases} ,
\label{IM_nD:3} \\
  = \frac{\bar{C}}{\pi^2 S}\frac{1+k'_{\lozenge}}{1-k'_{\lozenge}} \Bigg\{ 
          C^2\left[(1-\bar{S}^2){\rm E}_{\lozenge}-k'_{\lozenge}\bar{C}^2{\rm K}_{\lozenge}\right]
          \left( \frac{\bar{S}+S}{\bar{S}-S}\Pi_{\lozenge}+\frac{2\pi}{1+k'_{\lozenge}}\frac{\Theta(S-\bar{S})}{C\bar{C}}
          \right) \nonumber \\
           \phantom{\frac{\bar{C}}{\pi^2 S}\frac{1+k'_{\lozenge}}{1-k'_{\lozenge}} \Big\{}
          +\frac{4}{(1+k'_{\lozenge})^2}{\rm E}_{\lozenge}^2+k'_{\lozenge}(\bar{S}^2-S^2){\rm K}_{\lozenge}^2
          -(1-S^2)(1-\bar{S}^2){\rm E}_{\lozenge}{\rm K}_{\lozenge} \Bigg\} .
\label{IM_nD:4} 
\end{gather}

The correlation functions for the disorder variables or dual correlations are
given by
\begin{equation}
\langle \mu_{0,0}\mu_{N,N-1} \rangle
  = \det\begin{pmatrix}
          \tilde{a}_{0}  & \cdots & \tilde{a}_{-N+2} & \tilde{b}_{N-1} \cr
          \tilde{a}_{1} & \cdots & \tilde{a}_{-N+3} & \tilde{b}_{N-2} \cr
          \vdots & \vdots & \vdots  & \vdots  \cr
          \tilde{a}_{N-1} & \cdots & \tilde{a}_{1}  & \tilde{b}_{0}   \cr
        \end{pmatrix}, \; N \geq 1 ,
\end{equation}
where
\begin{equation}
 \tilde{b}_n = C\bar{S} \int_{\T}\frac{d\zeta}{2\pi i}\frac{\zeta^{n-1}}{\bar{S}+S\zeta}
                           \sqrt{\frac{1-k\zeta}{1-k/\zeta}} .
\end{equation}
The correlations in this sequence also have elliptic function evaluations analogous 
to (\ref{IM_nD:1}-\ref{IM_nD:4}) but we refrain from writing these down as they can 
be obtained from the direct correlations using the duality transformation
\begin{equation}
  \langle \mu_{0,0}\mu_{N,N-1} \rangle =
  \left.\langle \sigma_{0,0}\sigma_{N,N-1} \rangle
  \right|_{{\scriptstyle k\mapsto 1/k}\atop
          {{\scriptstyle S\mapsto 1/\bar{S}}\atop
           {\scriptstyle \bar{S}\mapsto 1/S}}} .
\end{equation}
In addition the $ \langle \sigma_{0,0}\sigma_{N-1,N} \rangle $ correlations can be 
obtained from $ \langle \sigma_{0,0}\sigma_{N,N-1} \rangle $ under the exchange 
$ S \leftrightarrow \bar{S} $.

These correlation functions are in fact characterised as a solution to an
{\em isomonodromic deformation} problem associated with the particular sixth 
Painlev\'e system, which itself characterises the diagonal correlation functions.
This observation is the key result of the present study.
\begin{proposition}\label{next-diag-Corr}
The next-to-diagonal correlation functions are given by the 
second type of associated functions (\ref{ops_eps:b}) appropriate to the weight 
(\ref{VI_wgt:a}) evaluated at a specific value of the spectral variable
\begin{equation}
  \langle \sigma_{0,0}\sigma_{N,N-1} \rangle
    = \frac{\bar{C}}{2\bar{S}}\frac{\DySt I_{N-1}}{\DySt \kappa_{N-1}}
      \epsilon_{N-1}^*(z=-\bar{S}/S) ,
\label{nextDiag}
\end{equation}
and valid for $ N \geq 1 $. Here $ I_{N} $ and $ \kappa_{N} $ are defined 
respectively by (\ref{ops_Uavge}) and (\ref{ops_I0}) appropriate to the weight (\ref{VI_wgt:a}).
\end{proposition}
\begin{proof}
A result in the general theory of bi-orthogonal polynomials is the determinantal 
representation with a Toeplitz structure for the reciprocal polynomial
\cite{FW_2004a}
\begin{equation}
  \phi^*_{n}(z) = \frac{\kappa_n}{I^{0}_{n}}
         \det \begin{pmatrix}
                     w_{0}      & \ldots & w_{-n+1}     & z^n  \cr
                     \vdots     & \vdots & \vdots       & \vdots \cr
                     w_{n-j}    & \ldots & w_{-j+1}     & z^j  \cr
                     \vdots     & \vdots & \vdots       & \vdots \cr
                     w_{n}      & \ldots & w_{1}        & 1  \cr
              \end{pmatrix} .   
  \label{ops_DetRep:b}
\end{equation}
Using this and the definition of second associated function (\ref{ops_eps:b}) one 
obtains an analogous bordered Toeplitz determinant \cite{Wi_2007}
\begin{equation}
  \epsilon^*_{n}(z) = \frac{\kappa_n}{I_{n}}
         \det \begin{pmatrix}
                     w_{0}      & \ldots & w_{-n+1}     & g_n  \cr
                     \vdots     & \vdots & \vdots       & \vdots \cr
                     w_{n-j}    & \ldots & w_{-j+1}     & g_j  \cr
                     \vdots     & \vdots & \vdots       & \vdots \cr
                     w_{n}      & \ldots & w_{1}        & g_0  \cr
              \end{pmatrix} ,
\label{eps_DetRep:b}
\end{equation}
where
\begin{equation}
   g_j(z) := -2z\int_{\T}\frac{d\zeta}{2\pi i\zeta}\frac{\zeta^j}{\zeta-z}w(\zeta),
   \quad z \notin \T . 
\end{equation}
The evaluation (\ref{nextDiag}) then follows by comparison of these last two 
formulae with (\ref{IM_nextDiag:a}) and (\ref{IM_nextDiag:b}).
\end{proof}

Many consequences flow from this identification - all of the general properties of 
the associated functions \cite{FW_2004a} can be applied. One particular useful
characterisation of the next-to-diagonal correlations is that they satisfy a
linear three-term recurrence relation. 
\begin{corollary}
The associated function (\ref{nextDiag}) satisfies the generic linear 
recurrence relation
\begin{equation}
  \frac{\kappa_{n}}{\kappa_{n+1}}\bar{r}_{n}\epsilon_{n+1}^*(z)
 +\frac{\kappa_{n-1}}{\kappa_{n}}\bar{r}_{n+1}z\epsilon_{n-1}^*(z)
 = [\bar{r}_{n}+\bar{r}_{n+1}z]\epsilon_{n}^*(z) ,
\label{TTrecur}
\end{equation}
subject to the two initial values for $ \epsilon^*_0, \epsilon^*_1 $
implied by (\ref{nextDiag}) and 
(\ref{IM_nD:1},\ref{IM_nD:2},\ref{IM_nD:3},\ref{IM_nD:4}).
The auxiliary quantities appearing in (\ref{TTrecur}) and (\ref{nextDiag})
satisfy the generic recurrences 
\begin{gather}
 I_{n+1} = \frac{I_{n}}{\kappa^2_{n}} ,
 \qquad
 \kappa_{n+1} = \frac{\kappa_{n}}{\sqrt{1-r_{n+1}\bar{r}_{n+1}}} ,
\end{gather}
subject to their initial values
\begin{equation}
 I_{0} = 1,\quad \kappa^2_{0} = \frac{1}{a_0} ,
\label{}
\end{equation}
utilising (\ref{ising_rRecur:c}).
\end{corollary}
We remark that this associated function also satisfies a linear second order differential
equation in the spectral variable $ z $ whose coefficients are determined by the auxiliary
quantities discussed above. However we refrain from writing this down as it doesn't appear
to have as much practical ultility as the recurrences in the above Corollary. 

To close our study we examine a number of limiting cases, namely the zero temperature, the 
critical temperature and high temperature limits.
At zero temperature, $ k \to \infty $, the solutions have leading order terms
$ (N \geq 1) $
\begin{equation}
     r_{N} \mathop{\sim}\limits_{k \to \infty} \frac{(-\half)_{N}}{N!}k^{-N}, 
     \quad
\bar{r}_{N} \mathop{\sim}\limits_{k \to \infty} \frac{(\half)_{N}}{N!}k^{-N} \; 
     \quad
   \langle \sigma_{0,0}\sigma_{N,N} \rangle \to 1 .
\end{equation}
At the critical point, $ k = 1 $, we have a complete solution for the bi-orthogonal 
system. The polynomial coefficients have the evaluations
\begin{equation}                                       
    \kappa^2_N = 
    \frac{\Gamma(N+\thalf)\Gamma(N+\half)}{\Gamma^2(N+1)}, \quad
     r_{N} = -\frac{1}{(2N+1)(2N-1)}, \quad
    \bar{r}_{N} = 1 ,
\end{equation}
which is consistent with the well known result \cite{McCW_1973} 
\begin{equation}                                       
    \langle \sigma_{0,0}\sigma_{N,N} \rangle = 
    \prod^{N}_{j=1}\frac{\Gamma^2(j)}{\Gamma(j+\half)\Gamma(j-\half)} .
\end{equation}
The isomonodromic system is 
\begin{align}
   \phi_N(z) & = -\frac{\kappa_N}{(2N+1)(2N-1)}\cdot{}_2F_1(\thalf,-N;-N+\thalf;z) ,
   \label{IM_crit:a} \\
   \phi^*_N(z) & = \kappa_N\cdot{}_2F_1(\half,-N;-N+\half;z) ,
   \label{IM_crit:b} \\
   \frac{1}{2}\kappa_N\epsilon_N(z) & 
     = -\frac{1}{(2N+3)(2N+1)z}\cdot{}_2F_1(\thalf,N+1;N+\fhalf;1/z) ,
   \label{IM_crit:c} \\
   \frac{1}{2}\kappa_N\epsilon^*_N(z) &
     = {}_2F_1(\half,N+1;N+\thalf;1/z) .
   \label{IM_crit:d}
\end{align}
This last result (\ref{IM_crit:d}) is consistent with the critical 
next-to-diagonal correlation given in \cite{A-YP_1987}
\begin{equation}                                       
    \langle \sigma_{0,0}\sigma_{N,N-1} \rangle = 
    \langle \sigma_{0,0}\sigma_{N,N} \rangle C \cdot{}_2F_1(\half,N;N+\half;-S^2) .
\end{equation}
At infinite temperature, $ k\to 0 $, the leading order terms are
$ (N \geq 1) $
\begin{equation}                                       
     r_{N} \mathop{\sim}\limits_{k \to 0} \frac{(-\half)_{N}}{(N+1)!}k^{-N},
     \quad 
\bar{r}_{N} \mathop{\sim}\limits_{k \to 0} \frac{N!}{(\half)_{N}}k^{N},
     \quad
    \langle \sigma_{0,0}\sigma_{N,N} \rangle \to  0 ,
\end{equation}
and the series expansion of these about $ k=0 $ in terms of the generalised 
hypergeometric function is given in \cite{FW_2004b}. 

This research has been supported by the Australian Research Council. The author would like
express his sincere gratitude for the generous assistance and guidance provided by Jacques Perk.
He has also benefited from extensive discussions on all matters relating to the Ising model 
in its various aspects with J.-M. Maillard, B. McCoy, T. Miwa and J. Palmer.  

\setcounter{equation}{0}
\bibliographystyle{amsplain}
\bibliography{moment,nonlinear,random_matrices}

\end{document}